\documentstyle[prb,aps,epsf,multicol,cite]{revtex}

\newcommand{\be}{\begin{equation}}
\newcommand{\ee}{\end{equation}}
\newcommand{\bea}{\begin{eqnarray}}
\newcommand{\eea}{\end{eqnarray}}
\newcommand{\nn}{\nonumber}

\newcommand{\rr}{{\bf r}}
\newcommand{\qq}{{\bf q}}
\newcommand{\kk}{{\bf k}}

\begin{document}
\flushbottom

\draft

\title{Fermi hypernetted-chain study of half-filled Landau levels with
	broken rotational symmetry }

\author{Orion Ciftja and Carlos Wexler}

\address{Department of Physics and Astronomy,
         University of Missouri--Columbia,
         Columbia, Missouri 65211, USA}

\date{\today}

\maketitle

\begin{abstract}
We investigate broken rotational symmetry (BRS) states at half-filling
of the valence Landau level (LL).  We generalize Rezayi-Read's (RR)
composite fermion (CF) trial wavefunctions to include anisotropic
coupling of the flux quanta to electrons, thus generating a nematic
order in the underlying CF liquid.  Using the Fermi hypernetted-chain
(FHNC) method which readily gives results in the thermodynamic limit,
we determine in detail the properties of these states.  By using the
anisotropic pair distribution and static structure functions we determine the
correlation energy and find that, as expected, RR's state is stable in
the lowest LL, whereas BRS states may occur at half-filling of higher LLs, 
with possible connection to the recently discovered quantum Hall liquid
crystals.

\end{abstract}
\vspace{-0.2cm}
%

\begin{multicols}{2}


\section{Introduction}
\label{sec:intro}
\vspace{-0.3cm}

During the past two decades the physics of two-dimensional electron
systems (2DES) has provided some of the richest grounds for new
developments in condensed matter physics.  In particular the integer
\cite{klitzing80} and fractional \cite{tsui82} quantum Hall effects
(QH) in strong magnetic fields are some of the most remarkable
phenomena discovered in the second half of the XX Century and rival
superconductivity in their fundamental significance by manifesting
quantum mechanics on macroscopic scales and providing a
major impetus to the development of new ideas in many-body physics
\cite{perspectives96}, like the existence of fractionally charged
quasiparticles  \cite{laughlin83}, topological quantum numbers
\cite{thouless98}, chiral Luttinger liquids \cite{wen91}, composite
particles \cite{composite}, etc.  One of the reasons that 2DES keep
supplying novel and exciting results is the improved quality of the
samples with mobilities increasing roughly exponentially with time,
thus allowing the emergence of new, subtler effects due to electronic
correlations (which are enhanced because of the reduced dimensionality).

Consider initially a 2DES at half-filling ($\nu = 1/2$) of the lowest
Landau level (LLL).  Experimentally, this state does not exhibit
the typical features of the fractional quantum Hall effect (FQHE),
namely the very precise quantization of the transverse conductance in
units of $e^2/h$, or the vanishing longitudinal resistivity.
Yet, the resistivity shows a broad minimum~\cite{willett1},
and anomalous behavior in the propagation of surface acoustic
waves \cite{willett2} has been observed.  Early
numerical work by Haldane~\cite{haldane}, suggested that $\nu = 1/2$
is not incompressible.  The overall experimental evidence strongly suggests
that, in the LLL, for half-filling the system behaves like a strange
Fermi liquid at close to zero {\em effective} magnetic field
\cite{onehalfexpt,sdh,rc}.

A theory of compressible Fermi-liquid-like behavior at half filling
has been proposed by Halperin, Lee, and Read (HLR) \cite{hlr}:
a two-dimensional (2D) system of electrons subjected to an external
perpendicular magnetic field, at half filling of the LLL can be
transformed to a mathematically  equivalent system of fermions
interacting with a Chern-Simons gauge field such that the average
effective magnetic field acting on the fermions is zero \cite{hlr}.
Since these fermions do not see a net magnetic field,
they can form a 2D Fermi sea of uniform density.  In a very successful
approach, based on the   composite fermion (CF) theory
\cite{composite,jain89}, Rezayi-Read \cite{RR}
(RR) described the half-filled state by a correlated Fermi
wave function that is a product of a Slater determinant of plane waves
with a Jastrow factor corresponding to a Bose Laughlin state at half
filling:
\bea
\label{hlr}
  \Psi (\rr_1, \ldots,\rr_N) = && \\
&&   \hspace{-2cm}
\hat{P}_{0} \; \prod_{j<k}^{N} (z_{j}-z_{k})^2 \;
	e^{-\sum_{k=1}^N |z_k|^2/4} \;
   \det \left[ \varphi_{\kk}(\rr_i)  \right]  \, , \nn
\eea
where $\varphi_{\kk}(\rr_i)$ are 2D plane waves
for fully spin polarized CF states that fill a 2D
disk in reciprocal space with Fermi momentum  $k_F$,
$z_j = x_j + i \, y_j$ is the complex 2D coordinate of $j$-th
electron and $\hat{P}_{0}$ is a projector onto the LLL ($L=0$).
We work in units of the magnetic length ($l_0^2=\hbar/eB=1$).

The situation is dramatically different for the next LL (LL index
$L=1$), where for $\nu \simeq 5/2$ a {\em quantized} Hall conductance
is observed along with a strong reduction of the longitudinal
conductance \cite{fivehalvesexpt}, which are an indication of a Cooper
pairing instability of the CFs \cite{52theory}.  A discussion of this
state is a wide and complex topic by itself and this, along with the
properties of the $\nu = 3/2$ and $7/2$ states (half filling of the
upper spin sub-band)  will not be addressed in this work.

In this paper we discuss some aspects related to the many interesting
new phenomena that have recently emerged in nearly half-filled higher
LLs (with LL index $L \ge 2$), in particular the extreme anisotropy
measured in the low temperature magnetotransport
\cite{lilly99a,du99,shayegan99}.  This anisotropic behavior has been
attributed to the formation of a nematic phase of the 2DES
which, at higher temperatures undergoes a nematic to isotropic
transition \cite{fradkin99,cw2000}.
The motivation of our work is to study these nematic phases by means of
many-body trial wavefunctions with broken rotation symmetry (BRS) at
half-filling of a Landau level.  Previously \cite{brsthird} we
reported on the existence of a BRS instability of the Laughlin state
\cite{laughlin83} for 1/3-filled higher LLs (with $L \ge 1$).  Here we
discuss a similar procedure for the more complex case of 1/2-filling,
which is more closely related to the experiments on anisotropic
phases \cite{lilly99a,du99,shayegan99}.

Similarly to the Musaelian and Joynt's (MJ) \cite{joynt} generalization
of Laughlin's FQHE state \cite{laughlin83} used by us in Ref.\
\onlinecite{brsthird}, we add a symmetry breaking parameter $\alpha$ in
the RR wavefunction [Eq.\ (\ref{hlr})]:

\bea
\label{eq:MJhalf}
\Psi_\alpha(\rr_1, \ldots ,\rr_N) =
\hat{P}_{L} \, \prod_{j<k}^{N} (z_j-z_k+\alpha)(z_j-z_k-\alpha)  && \nn \\
&& \hspace{-4.5cm}
	\times \; e^{-\sum_{k=1}^N |z_k|^2/4} \;
 {\det} \left[ \varphi_{\bf k}(\rr_i) \right]     \,,
\eea
where $\hat{P}_L$ now represents a projector onto the $L^{\rm th}$
LL.  This wavefunction represents a homogeneous Fermi-liquid-like
state at half filling factor,  and for $\alpha \neq 0$ has nematic
order (for $\alpha = 0$ we recover the RR wavefunction which is
obviously isotropic).  Note that the magnitude of $\alpha$ is related
to the amount of anisotropy and its phase to the angle the director of
the nematic (for  real $\alpha$ the system will have a stronger
modulation in the $x$-direction, and therefore likely have larger
conductance in the perpendicular direction: $\sigma_{yy} >
\sigma_{xx}$).  This wavefunction is an obvious starting point to
study the  {\em nematic} quantum Hall liquid crystals at half filling,
by facilitating the systematic study of the energy  dependence of BRS
states for diverse  physical parameters (LL index, width of the 2DES,
etc).


We study the BRS state for 1/2-filling of the valence LL
(i.e.\ $\nu = M+1/2$ with $M$ integer) by using the Fermi hypernetted-chain
(FHNC) method \cite{fanton1,krotscheck,zabo,manousakis,ristigclark}.
This method allows us to compute physical quantities in the
thermodynamic limit, without the limitations of using a finite number
of particles that hinder other techniques, where the extrapolation of
the results to the thermodynamic limit is not totally unambiguous.

We find that, for realistic interaction potentials, 
the RR (isotropic) state is stable in the LLL,
whereas a BRS state is possible in higher LLs.  While this is consistent
with the view of BRS states for $L \ge 2$, it does not reflect the
situation for $L=1$,
%
%
where it is believed that CF-s can form an incompressible state by
a Bardeen-Cooper-Schrieffer (BCS)-like pairing~\cite{bcs} as first pointed
out by Moore and Read~\cite{mr}.
Recent exact diagonalization results by Morf~\cite{morf} strongly suggest
that such spin-polarized Pfaffian state is the best candidate to
describe this filling.

In Sec.\ \ref{sec:hnc_theory} we present the basic theoretical
calculations needed to determine the stability of an isotropic or BRS
state.  A detailed description of the FHNC formalism in the context of
the BRS wave function [Eq.\ (\ref{eq:MJhalf})] is given in Sec.\
\ref{sec:hnc_method_BRS}.  The results for the BRS state in the LLL and
their extension in higher LLs are discussed in Sec.\
\ref{sec:results}.

\section{Basic theory}
\label{sec:hnc_theory}
\vspace{-0.3cm}

In this work we study the stability of different states by
using trial wavefunctions of the form of Eq.\ (\ref{eq:MJhalf}).  We
perform this analysis by comparing the energy in each of
these states to find the optimum value for the anisotropy-generating
parameter $\alpha$.  The potential, or correlation energy per
electron is given by:
\be
\label{potential}
	E_{\alpha} = \frac{1}{N}
	\frac{\langle \Psi_{\alpha}|\hat{V}| \Psi_{\alpha}\rangle}
	{\langle \Psi_{\alpha}|\Psi_{\alpha} \rangle}
= 	\frac{\rho}{2} \int d^2r \,  V(r) \,  \left[g(\rr) - 1\right] \,,
\ee
where $\hat{V}$ represents the electron-electron, electron-background, and
background-background interaction; and $g(\rr)$ is the
(angle-dependent) pair distribution function given by
\be
\label{eq:gr}
g(\rr) = \frac{N(N-1)}{\rho^2}
        \frac{ \int d^2r_3 \cdots d^2r_{N}
                | \Psi_\alpha (\rr_1 \cdots \rr_{N})|^2 }
        { \int d^2r_1 \cdots d^2r_{N}
                | \Psi_\alpha (\rr_1 \cdots \rr_{N})|^2 } \,,
\ee
where $\rr = \rr_2-\rr_1$.  The following normalization condition,
$\rho \int d^2r \left[ g(\rr)-1 \right] = -1$,
can be used as a convenient check for numerical procedures.
For an ideal 2D sample the interaction is a pure Coulomb potential
$V(r) \simeq e^2/(\epsilon r)$ , while in samples with finite thickness a
reasonable choice is the Zhang Das Sarma (ZDS) potential \cite{ZDS}
$V(r)=e^2/(\epsilon \sqrt{r^2+\lambda^2})$, where $\lambda$ is of the
order of the sample thickness.  Alternatively, the correlation energy
can be computed in reciprocal space:
\be
\label{eq:ener_sq}
E_\alpha = \frac{1}{2} \int \frac{d^2q}{(2\pi)^2} \, \tilde{V}(q) \,
\left[ S(\qq) - 1 \right] \,,
\ee
where $\tilde{V}(q)$ is the 2D Fourier transform \cite{ft} (FT) of
$V(r)$ and $S(\qq)$ is the static structure factor:
\be
\label{eq:sq}
S(\qq) - 1 =  \rho \, {\rm FT}[g(r) - 1] \,.
\ee

\begin{figure}
\begin{center}
\leavevmode
\epsfbox{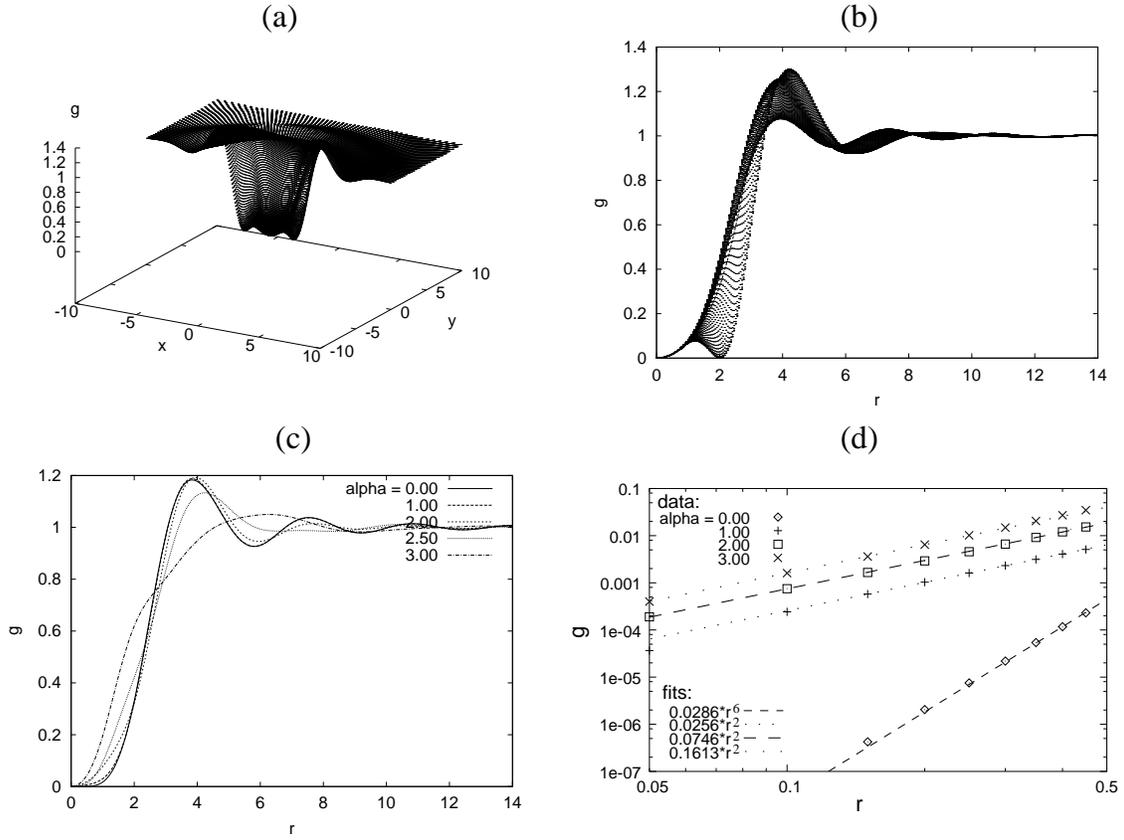}
\end{center}
\vspace{-0.cm}
\caption{ \label{fig:gr}
Pair distribution function for the BRS state at $\nu=1/2$.
{\em (a)} $\alpha=2$, surface plot of
$g(\rr)$ (the surface for $y<0$ was removed for clarity);
{\em (b)} $\alpha=2$, dotted lines: $g(r,\theta)$ for various
$\theta \in [0,2\pi]$, full line: angle averaged $\overline{g}(r)$;
{\em (c)} Angle averaged $\overline{g}(r)$ for
$\alpha$ = 0, 1, 2, 2.5 and 3;
{\em (d)} Small-$r$ behavior of $\overline{g}(r)$, lines are fitting
curves.
Note the discrete nodes of $g(r,\theta)$ at $r = \alpha$, $\theta =
\theta_\alpha, \theta_\alpha + \pi$ ($\theta_\alpha = 0$ in this
case). Calculations were performed in the FHNC/0 approximation.
}
\end{figure}

\begin{figure}
\begin{center}
\leavevmode
\epsfbox{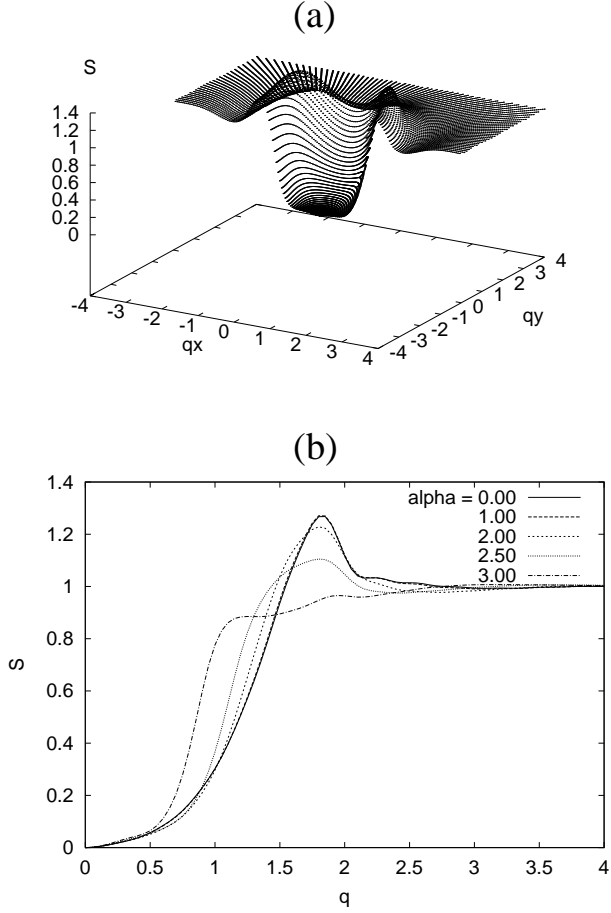}
\end{center}
\caption{ \label{fig:sq}
Static structure factor for the BRS state at $\nu=1/2$. {\em (a)}
$\alpha=2$, surface plot of $S(\qq)$ (the surface for $q_y<0$ was
removed for clarity);
{\em (b)} Angle averaged $\overline{S}(q)$ for
$\alpha$ = 0, 1, 2, 2.5 and 3.  Note the presence of peaks in $S(\qq)$
consistent with a nematic structure. Calculations were performed in
the FHNC/0 approximation.
}
\end{figure}

While both $g(\rr)$ and $S(\qq)$ are angle-dependent (e.g.\ see Figs.\
\ref{fig:gr} and \ref{fig:sq}), because the interaction potential is
centrally symmetric, the energy $E_\alpha$ depends only on the
angle-averaged pair distribution function or static structure factor
defined as:
\be
\label{angleav}
\overline{g}(r) = \int_{0}^{2 \pi} \frac{d \theta}{2 \pi} \; g(\rr) \,,
\hspace{0.5cm}
\overline{S}(q) =  \int_{0}^{2 \pi} \frac{d \theta_{q}}{2 \pi} \; S(\qq) \,.
\ee

The determination of either the pair distribution function or the
structure factor is generally a complicated integral problem that
needs to be solved for each LL.  However, its is known that if
transitions to other LLs are neglected (i.e.\ a {\em single-LL
approximation}), $g(\rr)$ and $S(\qq)$ at higher LL are simply related
to those at the LLL ($L=0$) by means of a convolution or product
respectively.  We will apply this approximation (which, moreover,
quenches the kinetic energy in higher LLs as well).  It is then,
sufficient to compute these distribution functions once in the LLL and
then the correlation energy per electron is given by
\be
\label{eq:ener_sq_N}
E^L_\alpha = \frac{1}{2} \int \frac{d^2q}{(2\pi)^2} \,
        \tilde{V}_{\rm eff}(q) \, \left[ S(\qq) - 1 \right] \,,
\ee
where $\tilde{V}_{\rm eff}(q) \equiv \tilde{V}(q) \, [L_L (q^2/2)]^2$.
$L_L(z)$ are Laguerre polynomials, and $S(\qq)$ is calculated in the
LLL ($L=0$).

In what follows we compute $g(\rr)$ and $S(\qq)$ using the FHNC method.

\vspace{-0.3cm}
\section{The Fermi HNC method for the broken rotational symmetry state}
\label{sec:hnc_method_BRS}
\vspace{-0.3cm}

The development of the FHNC method for Fermi systems
\cite{fanton1,krotscheck} allows one to
estimate accurately the expectation value of a Hamiltonian, the
pair distribution function and related quantities associated with a
Jastrow-Slater wave function and other more complex many-body
Fermi wave functions.
The FHNC method treats the correlated system of particles
{\em a priori} in the thermodynamic limit therefore
is extremely useful on the study of infinite homogeneous
Fermi systems described by correlated Fermi wave functions,
having (but not limited to) a Jastrow-Slater form.
In addition one can prove that the FHNC scheme achieves
convergence of the expectation value of a Hamiltonian within some expansion
scheme \cite{zabocon} thus substantiating the belief that FHNC gives
an accurate upper bound for the energy and other related quantities
\cite{zabocon,ciftjafantoni}.

An important problem arises at this stage.  The projection onto the
L$^{\rm th}$ LL performed in the RR wavefunction or its generalization
[Eq.\ (\ref{eq:MJhalf})] leads to a wavefunction that cannot be
directly treated within the FHNC formalism, because the simple
Jastrow $\times$ Slater determinant structure of single particle
orbitals is lost.  We therefore use an {\em unprojected} version of
Eq.\ (\ref{eq:MJhalf}), which although approximate, is believed to
contain the most important physics, especially since the Jastrow
factors already significantly annihilate higher-LL components of the
wavefunction \cite{trivedi}.  In addition, although the wave function
given in Eq.\ (\ref{eq:MJhalf}) has a Jastrow-Slater form after
dropping the projector $P_L$, the FHNC method in this case differs
from the standard approach since two-body correlation factors and
related quantities  depend not only on inter-particle distance, but
also on the relative angle between the particles.  In order to
calculate the pair distribution function the modulus square of the wave
function:

\bea
\label{modulus}
        |\Psi_{\alpha}(\rr_{1}, \ldots ,\rr_{N})|^2 =
	e^{\sum_{i>j}^{N} u(z_i - z_j)} \;
&&  \\
&& \hspace{-2.5cm} \times \;
	e^{-\sum_{i=1}^{N} \frac{|z_i|^2}{2}}
| {\rm det} \left[ \varphi_{\bf k}(\rr_i) \right]|^2    \,, \nn
\eea
where $u(z)=\ln |z-\alpha|^2+ \ln |z+\alpha|^2$,
is expanded perturbatively
in terms of $h(\rr_{ij})=\exp[{u(\rr_{ij})}]-1$ and may be ordered as:
%
\begin{eqnarray}
  |\Psi_{\alpha}|^2 &=&
 \left[\!1+\sum_{i<j}^{N} h(\rr_{ij})+\sum_{i<j}^{N} \sum_{k<l}^{N}
  h(\rr_{ij}) h(r_{kl})+ \cdots \! \right] \!  \nonumber \\ & &
\times
| {\rm det} \left[ \varphi_{\bf k}(\rr_i) \right]|^2   \ .
\label{eq4}
\end{eqnarray}
%
In addition to {\it dynamical correlations} between particles, described
by the pseudopotential $u(z_i-z_j)$, there are also
{\it statistical correlations} described by the Slater determinant which
renders the whole state antisymmetric.
Similar to the Jastrow part, we may expand
$| {\rm det} \left[ \varphi_{\bf k}(\rr_i) \right]|^2$ in the number
of exchanges between particles.
In this way Eq.\ (\ref{eq4}) becomes a quite symmetrical expansion in the
number of dynamical correlation factors, $h(\rr_{ij})$, as well as the number
of statistical correlation factors.
The resulting cluster terms in the expression for $g(\rr_{12})$
contain both kinds of correlations and may be represented by
cluster diagrams. As in the Bose case, the associated pair
distribution function $g(\rr_{12})$ is then given by the sum of all
linked irreducible diagrams obeying well-defined topological
rules \cite{fanton1}.

One defines {\it nodal}, {\it composite} (non-nodal), and {\it
  elementary} diagrams as in the Bose case, but there are now four
different types for each of them.  The four different classes of nodal,
composite
and elementary diagrams are generally denoted by {\it dd}
(direct-direct), {\it de} (direct-exchange), {\it ee}
(exchange-exchange), and {\it cc} (circular-exchange).
Then the pair distribution function is obtained from the
set of FHNC equations given in the Appendix, where $\rho$ is the density,
$g_s(=1)$ is the spin degeneracy and $l(r_{12})$ is the familiar statistical
exchange factor for the 2D Slater determinant given by
\begin{equation}
  l(\rr_1,\rr_2)= 2  \, \frac{J_{1}(k_{F} \ r_{12})}{k_{F} \ r_{12}}  \  ,
\label{eq20}
\end{equation}
where $r_{12}=|\rr_{2} - \rr_{1}| $ and $J_{1}(x)$ is the first
order Bessel function.

For standard systems like the $^3$He Fermi liquid,
the pair correlation factor is
short-range and heals to $1$ for large distances, therefore the
function $\exp[u(\rr_{ij})]-1$ provides a possible expansion parameter
[note that in order to apply the Fermi HNC expansion, the correlation
(pseudo) potential has to satisfy the conditions:
$u(\rr_{ij} \rightarrow 0) \rightarrow -\infty$ and
$u(\rr_{ij} \rightarrow +\infty) \rightarrow 0$].
In the case of the BRS wave function, the (pseudo) potential $u(\rr)$
is logarithmically long-range, however formally it is possible to extend the
method by splitting the pseudopotential associated with the Jastrow
part into
a short-- and a long--ranged part, respectively,

\begin{equation}
u(\rr_{12})= u_{s}(\rr_{12}) + u_{l}(\rr_{12}) \,,
\label{splitu}
\end{equation}
with the $dd$  nodal and composite function similarly split:
\bea
\label{split}
N_{dd}(\rr_{12}) &=& N_{dds}(\rr_{12}) - u_{l}(\rr_{12}) \,, \\
X_{dd}(\rr_{12}) &=& X_{dds}(\rr_{12}) + u_{l}(\rr_{12})  \,.
\eea
The splitting is done subject to the conditions:
\bea
\label{splitcond}
u(\rr_{12}) + N_{dd}(\rr_{12}) &=& u_s(\rr_{12}) + N_{dds}(\rr_{12}) \,, \\
N_{dd}(\rr_{12}) + X_{dd}(\rr_{12}) &=& N_{dds}(\rr_{12}) +
     X_{dds}(\rr_{12}) \,.
\eea
The short-range function $u_{s}(\rr_{12})$
(going to -$\infty$ for small distances and healing to $0$ for large
distances) and its its long-range counterpart are then chosen as:
\bea
\label{splitus}
u_{s}(\rr_{12}) &=&
- 2 \, K_0(Q |\rr_{12}\!-\!\vec{\alpha}|)
- 2 \, K_0(Q |\rr_{12}+\vec{\alpha}|)  \,, \\
u_{l}(\rr_{12}) &=& 
2 \, [ \ln(|\rr_{12}-\vec{\alpha}|) + K_0(Q |\rr_{12}-\vec{\alpha}|)] 
\nonumber \\
&& + 2 \, [ \ln(|\rr_{12}+\vec{\alpha}|)+ K_0(Q |\rr_{12}+\vec{\alpha}|)]
\,, 
\label{splitul}
\end{eqnarray}
where $K_0(x)$ is the modified Bessel function, and $Q$ is a cut-off
parameter of order 1.  We recall that the 2D FT\cite{ft} of
$u_{l}(\rr_{12})$ is:
\begin{equation}
\tilde{u}_{l}(\qq) = - \frac{4 \pi \, Q^2}{q^2(q^2+Q^2)}
\left( e^{i \qq \cdot \vec{\alpha}}
	+ e^{-i \qq \cdot \vec{\alpha}} \right) \,.
\label{ulq}
\end{equation}
The final set of equations is solved by a standard iterative procedure.  
There is one necessary approximation within the FHNC method in order
to obtain a
closed set of equations for the nodal and non-nodal functions: a small
set of cluster diagrams (corresponding to the so called {\em elementary}
diagrams), which cannot be included {\em ab initio} in the method,
needs to be somehow
estimated outside the FHNC method.  Several schemes have been devised
to include the contribution of such diagrams at various levels of
approximation, however the simplest approximation of totally neglecting
these terms (called the FHNC/0 approximation where we assume
$E_{dd}=E_{de}=E_{ee}=E_{cc}=0$) generally leads to very reliable
results \cite{zabo,zabocon,ciftjafantoni,ciftjanew} and we have adopted
it in this paper.

\vspace{-0.3cm}
\section{Results and discussions}
\label{sec:results}
\vspace{-0.3cm}

In this work we applied the FHNC theory to study the BRS state
at filling $1/2$ of an arbitrary LL.
For the sake of simplicity we neglected the elementary diagrams (FHNC/0).
This has allowed us to determine to a reasonable accuracy the pair
distribution function and the static structure factor.
In order to compare the $\alpha=0$ (RR state, {\em isotropic})
with the $\alpha \neq 0$ (BRS state, or {\em nematic}) we studied the
properties of the BRS wave function for several $\alpha$-s with
magnitudes between $0$ and $3$ (in general $\alpha = |\alpha| \, e^{i
\, \theta_\alpha}$, but without loosing generality we considered only
$\theta_\alpha = 0$).

%
%

In Fig.\ \ref{fig:gr} we plot the pair distribution function
$g(\rr)$ for $\alpha$ = 2 [panels (a) and (b)], and the angle-averaged
pair distribution function $\overline{g}(r)$ corresponding to $\alpha$
= 0, 1, 2, 2.5 and 3 [panels (c) and (d)].
It is interesting to note, for $\alpha \neq 0$,
the noticeable angle-dependence of $g(\rr)$, and the splitting of the
triple node at the origin to a simple node at the origin and
additional simple nodes at $r=\alpha$ and angle
$\theta=\theta_{\alpha}, \;\theta_\alpha+\pi$ ($\theta_\alpha=0$ in
this case).
From the energetics point of view [see Eq.\ (\ref{potential})], a
major consideration is the strong dependence of $\overline{g}(r)$  on
the value of the parameter $\alpha$.  As $\alpha$ is increased the
major peak of $\overline{g}(r)$ becomes less pronounced and for
$\alpha \approx 3$ it develops a shoulder for small $r$ [panel (c)].
In addition, note change in the small-$r$ behavior of
$\overline{g}(r)$ which switches from $\propto r^6$ (for $\alpha=0$)
to $\propto r^2$ as $\alpha$ is increased [panel (d)].
In general, for small-$r$, $\overline{g}(r)$  has almost no angular
dependence \cite{smallr}, and for $\alpha \neq 0$, $g(r \approx 0,
\theta) \simeq C_\alpha \, r^2$ for $0 \le r \le 0.5$,  where
$C_\alpha \simeq 0.024 \, \alpha^{1.7}$.


In Fig.\ \ref{fig:sq} we plot the static structure factor $S(\qq)$ for
$\alpha = 2$ (top panel), where the most important feature is the
emergence of peaks in $S(\qq)$ characteristic of a nematic structure;  
and the angle-averaged static structure factor $\overline{S}(q)$
corresponding to $\alpha$ = 0, 1, 2, 2.5 and 3 (bottom panel).  Note the
considerable dependence of $\overline{S}(q)$ on $\alpha$: as it increases
the peak is broadened and flattened, with no significant change in the
small-$q$ behavior.


One can compute the correlation energy per particle either directly
from Eqs.\ (\ref{potential}), (\ref{eq:ener_sq}), or
(\ref{eq:ener_sq_N}) to determine the energy per electron for
arbitrary values of the Fermi BRS parameter $\alpha$, the 2D system width
$\lambda$, and Landau level index $L$.  The following simplified
formula can be used in view of Eq.\ (\ref{angleav}):
\begin{equation}
E^L_{\alpha}(\lambda)= \frac{1}{4 \pi} \int_{0}^{\infty} \!\! dq \, q \,
\tilde{V}(q,\lambda) \, [L_L(\frac{q^2}{2})]^2 \,
	[ \overline{S}(q) - 1 ] \,,
\label{enerfroms}
\end{equation}
where
$\tilde{V}(q,\lambda) = ({2 \pi \, e^2}/{\epsilon \, q})
\, \exp(-\lambda q)$ is the 2D FT of the ZDS interaction
potential \cite{ZDS}.
In addition to allowing straightforward calculations to be extended to
any LL, Eq.\ (\ref{enerfroms}) permits higher numerical accuracy on
the calculation of $E_\alpha$ since $\overline{S}(q)$ saturates
exponentially to $1$ for relatively small values of $q$
as compared to $\overline{g}(r)$.

\begin{figure}
\begin{center}
\leavevmode
\epsfbox{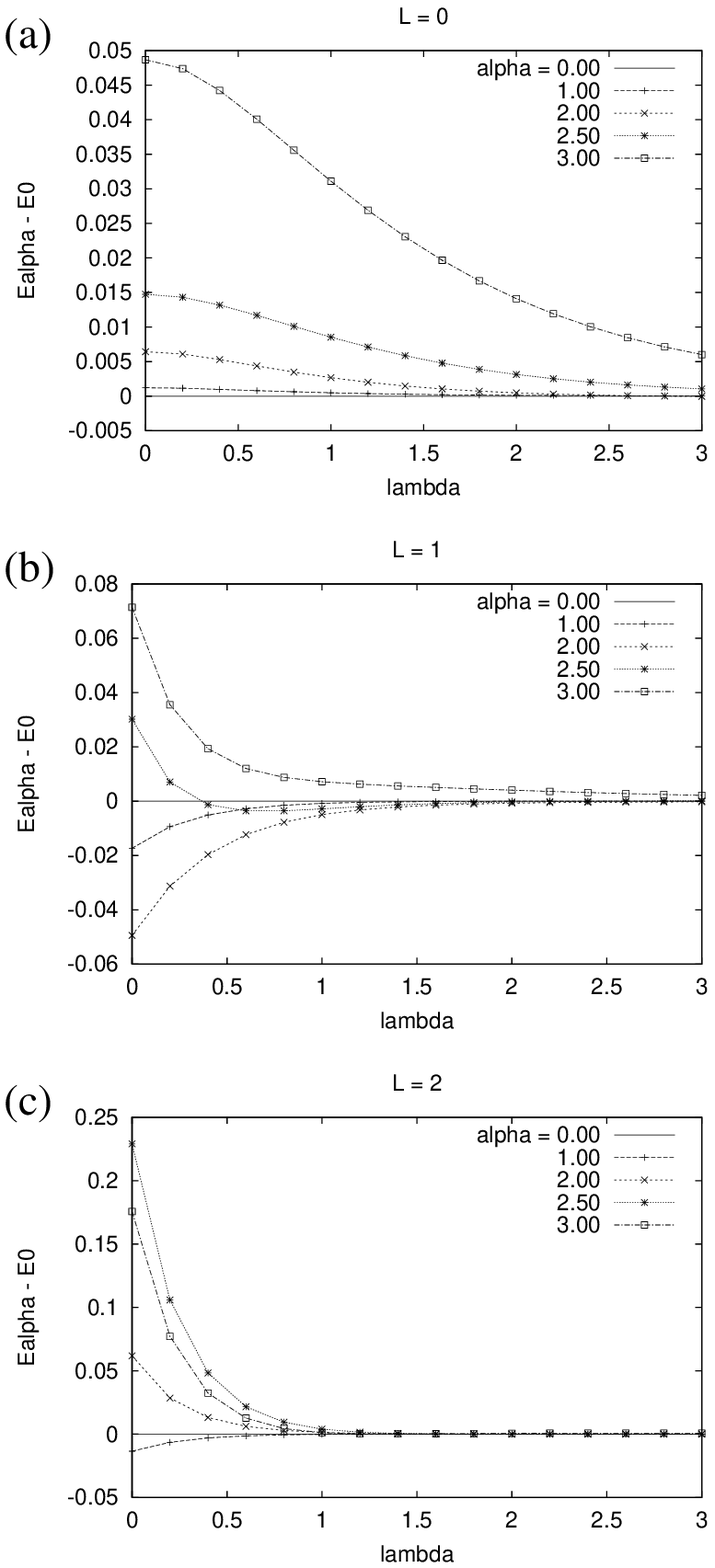}
\end{center}
\caption{  \label{fig:deltae}
	Energy per particle of the Fermi BRS states with
        $\alpha$ = 1, 2, 2.5 and 3
	relative to the isotropic ($\alpha=0$) state:
	$\Delta E_{\alpha}(\lambda) = E_{\alpha}(\lambda) -
	E_{0}(\lambda)$ for the LLL ($L=0$) and higher LL-s
        as function of 	the short distance cut-off parameter $\lambda$.
	Energies are in units of $e^2/(\epsilon l_o)$. Note that in
	the LLL, the Fermi BRS states always have an energy higher than
        the isotropic state,
	whereas in higher LLs ($L$ = 1, 2) there are ranges of
	$\lambda$ for which Fermi BRS states are favorable.
}
\end{figure}

Fig.\ \ref{fig:deltae} shows the energy difference between BRS states
with $\alpha$ = 1, 2, 2.5 and 3, and the isotropic state with $\alpha=0$.
Our findings indicate that in the LLL ($L=0$) the Rezayi-Read (RR) state is
stable for any $\lambda$, since all $\alpha \neq 0$ states have
higher energies (top panel).

The situation changes considerably in higher LLs ($L \ge 1$).  In
general, Fermi BRS states are found to have {\em lower} energies and
the compressible RR state is unstable towards a nematic state (see
lower panels of Fig.\ \ref{fig:deltae}).  Contrary to our findings for
1/3-filled LLs \cite{brsthird}, we do not see a runaway instability
but rather find that there are optimal values:  $\alpha_1^* \approx 2$
for the first LL ($L=1$), and  $\alpha_2^* \approx 1$ for $L=2$.
Whereas no BRS state has been so far observed 
for $L=1$
(likely because {\em another} correlated state not considered here is
lower in  energy) it is likely that the BRS state found here is
related to the low-temperature anisotropic conductance found in higher
LLs \cite{lilly99a,du99,shayegan99,fradkin99,cw2000}.


At this point it is important to comment on how precise our
determination of these energy differences is.  There are two aspects
to this problem.  First, the reader should note that we used an
{\em unprojected} wave function in place of the RR generalization
[Eq.\ (\ref{eq:MJhalf})].  While it would be highly desirable to
incorporate this LL projection operators into the formalism this is
not feasible within the FHNC method used here.  We should note,
however, that the presence of the Jastrow factors in Eq.\
(\ref{eq:MJhalf}), already 
provides a considerable projection into the LLL, 
and is believed to be particularly effective as
far as ground-state properties are concerned \cite{trivedi}.
The second aspect is that the FHNC/0 method is essentially a
variational method, giving energies that constitute an
upper bound to the exact ground state energy \cite{ripka}.  While a
precise estimate of these errors can only be made {\em a posteriori}
by comparing these results with alternative calculations, these
methods have proven to be quite reliable in scenarios ranging from
other partially filled LLs \cite{brsthird,ciftjafantoni} to $^3$He 
systems~\cite{zabo,zabocon}.  While the energy {\em differences} we are
interested (see Fig.\ \ref{fig:deltae}) are quite small, and perhaps
smaller than the errors in the absolute values of the correlation
energy in the states, we remark that these are not {\em uncorrelated
errors} but {\em systematic deviations} due to the nature of the
approximations used, and energy {\em differences} will likely be
considerably more precise.


In conclusion, we applied the FHNC theory to study possible Fermi BRS
states in a 1/2-filled LL.  We find that the isotropic RR state is
stable in the LLL for realistic interaction potentials.  In higher
LLs, BRS states with nematic order are energetically more favorable
than the RR state, perhaps with a direct connection to the anisotropic
states observed recently in high LLs \cite{lilly99a,du99,shayegan99}.


We would like to acknowledge helpful discussions with A.T.\ Dorsey,
M.\ Fogler, L.\ Radzihovsky and G.\ Vignale. This work was supported
by the University of Missouri Research Board.


\appendix
\vspace{-.3cm}
\section{}
\vspace{-.3cm}

For a Fermi system at density $\rho$ and spin degeneracy $g_s$ (1 or 2)
the sum of non-nodal (composite) diagrams
is given by:

\begin{equation}
X_{dd}(\rr_{12})=e^{u(\rr_{12})+N_{dd}(\rr_{12})
                     +E_{dd}(\rr_{12})}-N_{dd}(\rr_{12})-1
\label{A1}
\end{equation}

\begin{eqnarray}
X_{de}(\rr_{12}) &=& e^{u(\rr_{12})+N_{dd}(\rr_{12}) + E_{dd}(\rr_{12})}
                 \nonumber \\ & &
             \times  \left[N_{de}(\rr_{12})+E_{de}(\rr_{12})\right]
              -N_{de}(\rr_{12}) \ ,
\label{A2}
\end{eqnarray}

\begin{eqnarray}
X_{ee}(\rr_{12}) & = & e^{u(\rr_{12})+N_{dd}(\rr_{12})
                             +E_{dd}(\rr_{12})}
                       [ N_{ee}(\rr_{12})
                      \nonumber \\ & & \hspace{-1cm}
	+E_{ee}(\rr_{12})+
                   |N_{de}(\rr_{12})+E_{de}(\rr_{12})|^2
                    -g_{s} |N_{cc}(\rr_{12})
                 \nonumber \\ & & \hspace{-1cm}
                  +E_{cc}(\rr_{12})-l(\rr_{12})/g_{s}|^2 ]
                 -N_{ee}(\rr_{12}) \ ,
\label{A3}
\end{eqnarray}

\begin{eqnarray}
X_{cc}(\rr_{12}) &=& e^{u(\rr_{12})+N_{dd}(\rr_{12})
                          +E_{dd}(\rr_{12})}
                    [ N_{cc}(\rr_{12})+E_{cc}(\rr_{12})
                 \nonumber \\ & &
                  -l(\rr_{12})/g_{s}] +l(\rr_{12})/g_{s}
                  -N_{cc}(\rr_{12}) \ .
\label{A4}
\end{eqnarray}

The chain formation of the nodal diagrams is generated by convolution
equations

\begin{equation}
N_{dd}(\rr_{12})= \rho \int d \rr_{3}
  [X_{dd}(\rr_{13})+N_{dd}(\rr_{13})] P(\rr_{32}) \ ,
\label{A5}
\end{equation}

\begin{eqnarray}
\label{A6}
N_{de}(\rr_{12}) &=& \rho \int d \rr_{3}
                     [X_{dd}(\rr_{13}) X_{ee}(\rr_{32})
                 \\ & & \hspace{-1cm}
                     -X_{de}(\rr_{13}) X_{de}(\rr_{32})
             +[X_{de}(\rr_{13})+N_{de}(\rr_{13})] P(\rr_{32})] , \nn
\end{eqnarray}

\begin{eqnarray}
\label{A7}
N_{ee}(\rr_{12}) &=& \rho \int d \rr_{3} [X_{de}(\rr_{13})
               X_{de}(\rr_{32})
                 \\ & & \hspace{-2cm}
	- X_{dd}(\rr_{13}) X_{ee}(\rr_{32})
            +[X_{ee}(\rr_{13})+N_{ee}(\rr_{13})] P(\rr_{32})] \ , \nn
\end{eqnarray}

\begin{eqnarray}
N_{cc}(\rr_{12}) &=& \rho \int d \rr_{3} [-l(\rr_{13})/g_{s}
                          +X_{cc}(\rr_{13})
                 \nonumber \\ & &
                  +N_{cc}(\rr_{13})] X_{cc}(\rr_{32}) \ ,
\label{A8}
\end{eqnarray}

where
\begin{eqnarray}
\label{A9}
P(\rr_{ij}) &=& X_{dd}(\rr_{ij})+2 X_{de}(\rr_{ij})
                  \\ & &  \hspace{-2cm}
                  +\rho \int d \rr_{k}
                [X_{dd}(\rr_{ik}) X_{ee}(\rr_{kj})
                 -X_{de}(\rr_{ik}) X_{de}(\rr_{kj})]  . \nn
\end{eqnarray}
The pair distribution function is then given from:
\begin{eqnarray}
\label{eq11}
  g(\rr_{12}) &=& 1+X_{dd}(\rr_{12})+N_{dd}(\rr{12})
                                     \\ & & \hspace{-1cm}
	+2 \ [X_{de}(\rr_{12})+N_{de}(\rr_{12})]
                 +X_{ee}(\rr_{12})+N_{ee}(\rr_{12}) \ . \nn
\end{eqnarray}

\end{multicols}

\noindent \hrule
\medskip
\newpage
\centerline{{\bf FIGURES}}

\newpage
\begin{multicols}{2}

\end{multicols}

\vspace{-0.3cm}
\references
\vspace{-1.6cm}



\bibitem{klitzing80}
        K.\ von Klitzing, G.\ Dorda and M.\ Pepper,
        Phys.\ Rev.\ Lett.\ {\bf 45}, 494 (1980).

\bibitem{tsui82}
        D.C.\ Tsui, H.L.\ Stormer and  A.C.\ Gossard,
        Phys.\ Rev.\ Lett.\ {\bf 48}, 1559 (1982).

\bibitem{perspectives96}
        {\em Perspectives in quantum Hall effects,}
        ed.\ by  S.\ Das Sarma and A.\ Pinczuk (Wiley, New York 1996).

\bibitem{laughlin83}
        R.B.\ Laughlin,
        Phys.\ Rev.\ Lett.\ {\bf 50}, 1395 (1983).

\bibitem{thouless98}
        D.J.\ Thouless,
        {\em Topological quantum numbers in nonrelativistic physics,}
        (World Scientific, Singapore, 1998).

\bibitem{wen91}
        X.G.\ Wen,
        Phys.\ Rev.\ B {\bf 43}, 11025 (1991);
        Phys.\ Rev.\ Lett.\ {\bf 64}, 2206 (1990).

\bibitem{composite}
       	For a review see: J.\ Jain,
       	{\em The composite fermion: a quantum
       	particle and its quantum fluids,}
       	Physics Today, Apr/2000, p.\ 39; and/or
	{\em Composite Fermions}, ed.\ O.\ Heinonen
        (World Scientific, New York, 1998).

\bibitem{willett1}
	R.L.\ Willett, J.P.\ Eisenstein, H.L.\ Stormer,
 	D.C.\ Tsui, A.C.\ Gossard, and J.H.\ English,
	Phys.\ Rev.\ Lett.\ {\bf 59}, 1776 (1987).

\bibitem{willett2}
	R.L.\ Willett, M.A.\ Paalanen, R.R.\ Ruel,
        K.W.\ West, L.N.\ Pfeiffer, and D.J.\ Bishop,
 	Phys.\ Rev.\ Lett.\ {\bf 65}, 112 (1990).

\bibitem{haldane}
	F.D.M.\ Haldane,
	Phys.\ Rev.\ Lett.\ {\bf 55}, 2095 (1985).


\bibitem{onehalfexpt}
        For a review of experimental evidence for filling factor
        $\nu = 1/2$ see H.L.\ Stormer and D.C.\ Tsui,
        {\em Composite Fermions in the Fractional Quantum Hall Effect},
        in Ref.\ \protect\cite{perspectives96}.

\bibitem{sdh}
        R.R.\ Du, H.L.\ Stormer, D.C.\ Tsui, L.N.\ Pfeiffer, and
        K.W.\ West,
        Solid State Comm. {\bf 90}, 71 (1994).

\bibitem{rc}
        V.J.\ Goldman, B.\ Su, and J.K.\ Jain,
        Phys.\ Rev.\ Lett.\ {\bf 72}, 2065 (1994).

\bibitem{hlr}
	B.I.\ Halperin, P.A.\ Lee, and N.\ Read,
	Phys.\ Rev.\ B {\bf 47}, 7312 (1993).

\bibitem{jain89}
	 J.K.\ Jain,
	Phys.\ Rev.\ Lett.\ {\bf 63}, 199 (1989).

\bibitem{RR}
	E. Rezayi, and N. Read,
        Phys. Rev. Lett. {\bf 72}, 900 (1994).

\bibitem{fivehalvesexpt}
        R.L.\ Willett,
        J.P.\ Eisenstein, H.L.\ Stormer, D.C.\ Tsui, A.C.\ Gossard,
        J.H.\ English,
        Phys.\ Rev.\ Lett.\ {\bf 59}, 1779 (1987);
        W.\ Pan,
        J.-S.\ Xia, V.\ Shvarts,D.E.\ Adams,H.L.\ Stormer,D.C.\ Tsui,
        L.N.\ Pfeiffer, K.W.\ Baldwin, and K.W.\ West
        Phys.\ Rev.\ Lett.\ {\bf 83}, 3530 (1999).

\bibitem{52theory}
	See e.g.:
        V.W.\ Scarola, K.\ Park, and J.K.\ Jain,
        Nature {\bf 406}, 863 (2000),
	and references therein;
        N.\ Read,
        preprint cond-mat/0010071, unpublished;
        N.\ Read,
	Physica B {\bf 298}, 121 (2001), and reference therein.

\bibitem{lilly99a}
        M.P.\ Lilly,
        K.B.\ Cooper, J.P.\ Eisenstein,
        L.N.\ Pfeiffer, and K.W.\ West,
        Phys.\ Rev.\ Lett.\ {\bf 82}, 394 (1999).

\bibitem{du99}
        R.R.\ Du,
        D.C.\ Tsui, H.L.\ Stormer, L.N.\ Pfeiffer,
        K.W.\ Baldwin, and K.W.\ West,
        Solid State Comm.\ {\bf 109}, 389 (1999).

\bibitem{shayegan99}
        M.\ Shayegan, H.C.\ Manoharan, S.J.\ Papadakis,
        E.P.\ DePoortere,
        Physica E {\bf 6}, 40 (2000).



\bibitem{fradkin99}
        E.\ Fradkin and S.A.\ Kivelson,
        Phys.\ Rev.\ B {\bf 59}, 8065 (1999).


\bibitem{cw2000}
        C.\ Wexler and A.T.\ Dorsey,
        Phys.\ Rev.\ B {\bf 64}, 115312 (2001).

\bibitem{brsthird}
	O.\ Ciftja and C.\ Wexler,
	to appear in Phys.\ Rev.\ B (15/Jan/2002),
	preprint cond-mat/0108119.

\bibitem{joynt}
	K.\ Musaelian and R.\ Joynt,
	J.\ Phys.: Condens.\ Matter {\bf 8}, L105 (1996).


%




\bibitem{fanton1}
	S.\ Fantoni and S.\ Rosati,
	Nuovo Cim.\ Lett.\ {\bf 10} 545 (1974);
	S.\ Fantoni and S.\ Rosati,
	Nuovo Cim.\ {\bf A 25} 593 (1975).

\bibitem{krotscheck}
	E. Krotscheck and M.L. Ristig, Phys. Lett. {\bf 48A}, 17 (1974);
	E. Krotscheck and M.L. Ristig, Nucl. Phys. A {\bf 242}, 389 (1975).

\bibitem{zabo}
	J.G. Zabolitzky, Phys. Rev. A {\bf 16}, 1258 (1977).

\bibitem{manousakis}
        E. Manousakis, S. Fantoni, V.R. Pandharipande, and Q.N. Usmani,
	Phys. Rev. B {bf 28}, 3770 (1983).

\bibitem{ristigclark}
	M.L. Ristig and J.W. Clark,
	Phys. Rev. B {bf 14}, 2875 (1976).


\bibitem{bcs}
	J. Bardeen, L.N. Cooper, and J.R. Schrieffer,
	Phys. Rev. {bf 106}, 162 (1957).
\bibitem{mr}
	G. Moore and N. Read,
	Nucl. Phys. B {bf 360}, 362 (1991).
\bibitem{morf}
	R.H. Morf,
        Phys.\ Rev.\ Lett.\ {\bf 80}, 1505 (1998).



\bibitem{ZDS}
	F.C.\ Zhang and S.\ Das Sarma,
	Phys.\ Rev.\ B {\bf 33}, 2903 (1986).

\bibitem{ft}
	We use the standard convention for the 2D FT:
	$\tilde{f} (\qq) = \int d^2r \, \exp[-i \qq \cdot \rr] \,
	f(\rr)$,
	$f(\rr) = \int d^2q/(2 \pi)^2 \,  \exp[i \qq \cdot \rr] \,
	\tilde{f} (\qq)$.

\bibitem{zabocon}
	J.G. Zabolitzky, Phys. Lett. {\bf 64B}, 233, (1976).


\bibitem{ciftjafantoni}
	O.\ Ciftja and S.\ Fantoni,
        Phys.\ Rev.\ B {\bf 56}, 13290 (1997).

\bibitem{trivedi}
	N.\ Trivedi, J.K.\ Jain,
	Mod.\ Phys.\ Lett.\ B {\bf 5}, 503 (1993).



\bibitem{ciftjanew}
	O.\ Ciftja, S.\ Fantoni, J.W.\ Kim and M.L.\ Ristig,
        J. Low Temp. Phys. {\bf 108}, 357 (1997).


\bibitem{smallr}
The absence of angular dependence on $g(r,\theta)$ for small
$r$ can be easily understood by noting that in the small-$r$
limit: 
$g(r \approx 0,\theta) \propto \exp[u_{s}(r \approx 0,\theta)] \
[ 1- l(r \approx 0)^2/g_s ]$, ($g_s=1$) 
where $ u_{s}(r, \theta)$ is given from Eq.\
(\ref{splitus}). The second term in the above expression
is angle-independent and represents the zero of the Pauli principle.
For short distances: 
$\lim_{r \rightarrow 0} K_{0}(Q r)=-\ln({Q r}/{2})-\gamma$, 
($\gamma=0.5772...$ is Euler's constant), therefore
$u_{s}(r \approx 0,\theta)]$ and by consequence
$g(r \approx 0,\theta)$ is angle-independent in this region.
For more details see Ref.\ \protect\onlinecite{brsthird}.

\bibitem{ripka}
	G.\ Ripka,
	Physics Reports {\bf 56}, 1 (1979).
\end{document}